\documentclass[letterpaper,twocolumn,10pt]{article}
\usepackage{usenix}

\newcommand{\sssec}[1]{\noindent\textbf{#1} }
\usepackage{amssymb}
\usepackage{xspace}
\usepackage{amsmath}
\newcommand{\sysname}{\textsc{UiAs}\xspace}

\usepackage{marvosym} 

\usepackage{graphicx}
\usepackage{subcaption}
\usepackage{float}
\usepackage{booktabs}
\usepackage{multirow}
\usepackage{makecell}
\begin{document}

\date{}

\title{\Large \bf UiAs: User-Independent 3D Facial Anti-Spoofing via Multi-modal Wireless Signals}

\author{
Zhiwei Chen$^{1}$ \quad
Lebin Lyu$^{1}$ \quad
Yimo Zhang$^{1}$ \quad
Dingyu Zhong$^{1}$ \quad
Yijie Li$^{2}$\\[3pt]
Yichao Chen$^{3}$ \quad
Dian Ding$^{3}$ \quad
Jiguo Yu$^{1}$ \quad
Xiaosong Zhang$^{1}$ \quad
Yongzhao Zhang$^{1}$\raisebox{0.7ex}{\scriptsize\Letter}\\[7pt]
\textit{$^{1}$University of Electronic Science and Technology of China}\\
\textit{$^{2}$Shanghai University of Finance and Economics}\\
\textit{$^{3}$Shanghai Jiao Tong University}\\[3pt]
\Letter\ \textit{Corresponding Author}
}

\maketitle


\begin{abstract}
Face authentication is widely deployed in security-sensitive applications, while increasingly realistic 3D spoofing attacks pose growing threats. High-fidelity 3D masks can reproduce facial appearance and geometry but cannot replicate the intrinsic physical responses of living tissue, which can be actively probed by wireless signals. However, the resulting liveness cues captured by wireless signals are entangled with user-dependent facial geometry, limiting cross-user generalization. We present \sysname, a multimodal user-independent 3D facial anti-spoofing system using electromagnetic (mmWave) and mechanical (acoustic) waves. The two modalities share similar user-dependent geometric variations, allowing \sysname to suppress them through cross-modal subtraction while preserving modality-specific liveness cues. Their complementary physical responses further improve live/spoof discrimination. In practical deployments, multiple materials (e.g., skin, hair, eyeglasses, or face coverings) may also bias liveness representations, while spoofing materials are diverse and open-ended. \sysname addresses both through skin-anchored contrastive learning. We evaluate \sysname with real 3D spoofing attacks, which achieves 93.25\% accuracy for unseen users without user-specific physical-signal enrollment.
\end{abstract}

\section{Introduction}
\label{sec:intro}
Face authentication has been widely deployed in mobile devices, access-control systems, financial services, and identity-verification platforms~\cite{background}. Its growing adoption has also made it an attractive target for face spoofing attacks, in which an adversary presents a fabricated representation of a legitimate user to the sensing device. Successful attacks may result in account takeover~\cite{accounttakeover}, fraudulent transactions~\cite{fraudulenttransactions}, privacy breaches~\cite{privacy}, and unauthorized access~\cite{mobilebanking,accesscontrol} to digital services or physical facilities. Recent real-world incidents further demonstrate the severe financial consequences of identity-related attacks~\cite{identitytheft,3dlossmoney}. These attacks undermine the reliability and security of facial authentication systems.
\begin{figure}[t]
    \centering

    \begin{minipage}[c]{0.50\linewidth}
        \centering

        \begin{subfigure}{\linewidth}
            \centering
            \includegraphics[width=\linewidth]{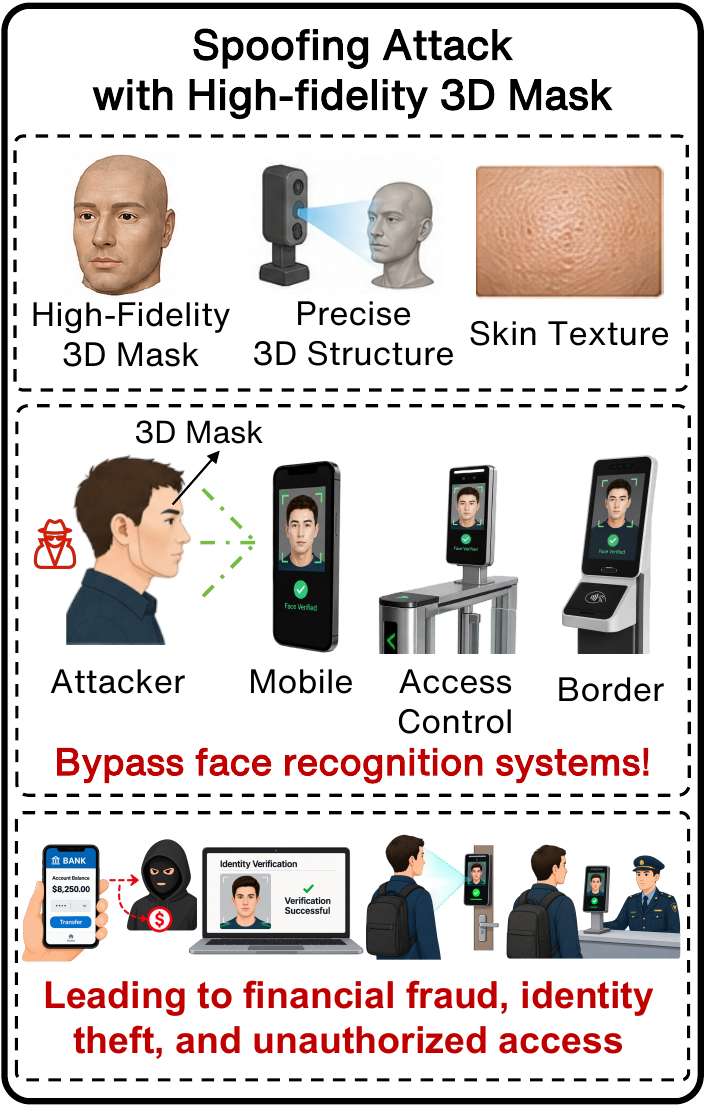}
            \caption{High-fidelity 3D spoofing attacks pose severe security threats to face authentication systems.}
            \label{fig:intro1}
        \end{subfigure}

    \end{minipage}
    \hfill
    \begin{minipage}[c]{0.49\linewidth}
        \centering

        \begin{subfigure}{\linewidth}
            \centering
            \includegraphics[width=\linewidth]{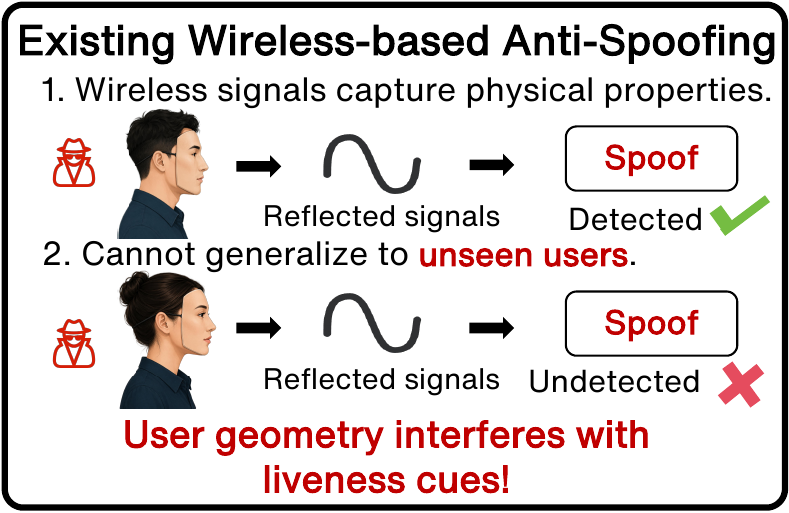}
            \caption{Single-modal probing captures liveness cues but remains user-dependent.}
            \label{fig:intro2}
        \end{subfigure}

        \par

        \begin{subfigure}{\linewidth}
            \centering
            \includegraphics[width=\linewidth]{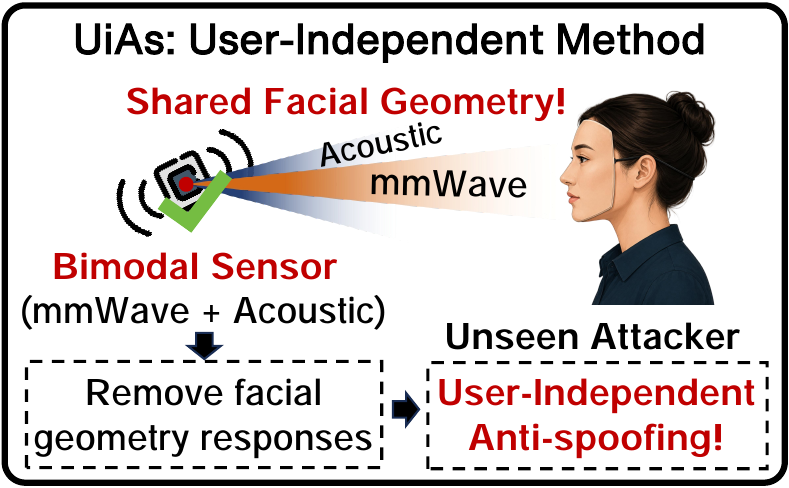}
            \caption{\sysname suppresses nuisance information for user-independent anti-spoofing.}
            \label{fig:intro3}
        \end{subfigure}

    \end{minipage}

    \caption{Illustration of \sysname. \sysname combines electronic and mechanical waves to suppress shared user-dependent variations and preserve liveness cues for user-independent 3D face anti-spoofing.}
    \label{fig:intro}

\end{figure}

Face spoofing attacks are evolving from conventional 2D presentation media toward high-fidelity 3D masks. Early 2D attacks primarily relied on planar media, such as printed photographs~\cite{cvprinted} and screen replays. Accordingly, face anti-spoofing methods detect visual inconsistencies using texture~\cite{cvtexture} or motion~\cite{cvmotion1,cvmotion2,cvmotion3}, while depth~\cite{cvdepth1,cvdepth2} and thermal~\cite{rgbthermal} sensing provide additional structural and temperature information. However, advances in facial scanning, molding, and 3D printing have enabled high-fidelity 3D spoofing masks that reproduce realistic skin appearance and facial geometry~\cite{cvmask}, move coherently with the attacker once worn, and absorb body heat, thereby weakening defenses based on texture, motion, depth, and temperature.

To overcome the limitations of passive observation, existing methods employ active electromagnetic probing~\cite{mmface,rface} or acoustic probing~\cite{echoface,beyond,sonarguard,aface,ultraface} to capture physical-response differences between spoofing materials and living facial tissue. However, the resulting liveness cues (i.e., physical-response evidence used to distinguish live faces from spoofing attacks) remain deeply entangled with user-dependent facial geometry, biasing representations toward identities observed during training and undermining cross-user generalization and deployment scalability. Existing systems are typically designed for user-dependent authentication or template-based verification and compare sensed facial responses against enrolled user-specific templates, as illustrated in Fig.~\ref{fig:intro2}. Such designs generalize poorly to unseen users (i.e., target identities impersonated by 3D masks but not previously enrolled in the system) and require per-user physical-signal enrollment, which is impractical in open-world deployments such as public access control and border screening. This raises a fundamental question: \textbf{how can high-fidelity 3D mask attacks be reliably detected for unseen users without user-specific physical-signal enrollment?}

In this work, we present \sysname, a user-independent 3D face anti-spoofing system that combines electromagnetic (mmWave) and mechanical (acoustic) waves. Our key observation has two aspects: (1) when the two sensors are co-located, they probe the same face from nearly the same viewpoint and therefore encode similar variations induced by user-dependent facial geometry, as illustrated in Fig.~\ref{fig:intro3}; and (2) the two modalities characterize different physical properties of the presented face, such as dielectric properties (electromagnetic) and acoustic impedance (mechanical). Together, these properties allow \sysname to suppress shared user-dependent geometric variations while preserving complementary material-dependent cues for user-independent liveness detection.
Specifically, \sysname addresses two technical challenges:

\textit{1) How to reliably disentangle user-independent liveness cues from dominant geometry-related responses.}
Facial geometry typically exerts a much stronger influence on the sensed signals than the subtle physical differences introduced by the presented medium. Consequently, mmWave and acoustic representations jointly encode dominant user-dependent geometric responses together with weaker liveness-related responses associated with properties such as complex permittivity and acoustic impedance. This imbalance can cause conventional correlation-based disentanglement to be dominated by shared geometry and overlook liveness-discriminative information. To address this problem, \sysname explicitly suppresses the dominant shared component through cross-modal subtraction in the latent space. To enable meaningful cross-modal comparison, it first aligns the heterogeneous representations and calibrates their scales, making corresponding components directly comparable. It then subtracts the calibrated representations to suppress shared geometry-related variations while preserving modality-specific liveness responses. To further disentangle the retained liveness information from residual geometry-related factors, \sysname applies orthogonality and reconstruction constraints, reducing information overlap while preventing feature collapse. Together, these operations extract liveness cues that generalize to unseen users.

\textit{2) How to reliably project liveness cues under the material variations.}
In practical deployments, the sensed facial region may contain multiple materials, such as genuine skin or a spoofing medium together with hair, eyeglasses, or face coverings. Since different materials exhibit distinct electromagnetic and mechanical responses, they can distort the observed physical response and bias the extracted liveness representation. Such interference cannot be removed by cross-modal subtraction because it appears as modality-specific responses rather than shared geometry-related information. Moreover, spoofing attacks may employ diverse and evolving materials, making it impractical to explicitly model every possible spoofing medium. \sysname therefore focuses on stable physical characteristics associated with genuine skin rather than individual spoofing materials. To achieve this, \sysname introduces skin-anchored contrastive projection that maps material-affected representations into a common liveness space. Genuine samples are pulled toward a clean-skin anchor, while spoofing samples serve as negatives. Liveness is then determined by the consistency between the projected representation and the skin anchor. This objective suppresses material-induced feature drift while avoiding exhaustive modeling of spoofing materials.

We implement \sysname using commercial off-the-shelf (COTS) mmWave and acoustic sensors and evaluate it under variations in user identity, 3D spoofing media, and multi-material interference. The results show that \sysname can preserve liveness-discriminative physical evidence while reducing user-dependent interference, enabling liveness detection without user-specific physical-signal enrollment.

This work makes the following contributions:
\begin{itemize}
\item To the best of our knowledge, \sysname is the first user-independent wireless face anti-spoofing system to detect high-fidelity 3D spoofing attacks.

\item We design a cross-modal user-independent feature extraction framework that suppresses facial geometry variations while preserving modality-specific liveness cues.

\item We develop a skin-anchored contrastive liveness projection module that mitigates multi-material interference and handles open-ended spoofing materials by aligning representations with clean genuine-skin anchors.

\item We evaluate \sysname 
with real 3D spoofing masks and multi-material conditions. \sysname achieves 93.25\% accuracy and consistently outperforms single-modal and conventional multimodal baselines under unseen-user evaluations.
\end{itemize}

\section{Background}
\label{sec:background}
\subsection{High-Fidelity 3D Facial Spoofing Attack}
\label{subsec:3d_spoofing}
Face spoofing attacks have evolved from printed photographs and screen replays~\cite{cvprinted} to high-fidelity 3D artifacts that reproduce both the appearance and facial structure of a target user. Meanwhile, fabricating such customized 3D masks has become increasingly accessible. While earlier attacks required dedicated 3D scanning and precise facial modeling, modern pipelines can reconstruct facial geometry from multi-view images readily obtained from social media and fabricate customized masks through printing, molding, or casting~\cite{JIA2020107032}. This increased accessibility is accompanied by substantial improvements in fabrication fidelity: recent systems achieve a manufacturing accuracy of approximately 1.66~mm~\cite{3Dreproduce}, while coloring and texture processing further approximate the appearance of genuine skin. These masks can also be fabricated from diverse materials, including resin, latex, and silicone~\cite{cvmask} and etc., making it difficult to explicitly model every possible spoofing medium for liveness detection.

High-fidelity 3D mask attacks have already been reported in security-critical applications, including mobile banking~\cite{mobilebanking}, remote identity verification~\cite{remoteidentity}, access control~\cite{accesscontrol}, and border inspection~\cite{border}. 
As early as 2010, in the U.S., a bank robber used a realistic 3D mask to impersonate another individual, leading to the wrongful arrest of an innocent person~\cite{3dlossmoney2};
more recently, a fraudster reportedly used a high-fidelity 3D mask to impersonate the French Minister of Defence and caused approximately $90$ million in losses~\cite{3dlossmoney}. These incidents highlight the growing practical threat posed by high-fidelity 3D facial impersonation.

\sssec{$\blacksquare$\textbf{Key observation:}}
Regardless of their fabrication precision or constituent material, 3D spoofing masks cannot reproduce the intrinsic physical responses of living facial tissue. Detecting genuine skin responses therefore provides a more generalizable defense against diverse high-fidelity 3D attacks.

\subsection{Active Probing with Wireless Signals}
\label{subsec:skin_detection}
Although the intrinsic physical responses of genuine skin provide a robust basis for detecting high-fidelity 3D masks, conventional visual sensing cannot directly capture them. RGB cameras mainly observe appearance information such as color and texture~\cite{cvtexture}, while depth and thermal cameras provide additional geometric and temperature information~\cite{cvdepth1,cvdepth2,rgbthermal}. However, high-fidelity masks can closely reproduce facial appearance and contours and may absorb body heat after being worn, weakening these cues. Therefore, visual sensing remains insufficient for characterizing the intrinsic physical properties of the presented medium.

In contrast, active probing with wireless signals can directly characterize the physical responses of the presented medium by transmitting known signals and analyzing the reflected waveforms. Electromagnetic probing can be implemented using radio-frequency (RF) technologies, including WiFi~\cite{intuwition}, UWB~\cite{siwa}, mmWave, and RFID~\cite{rface,mmface}, while mechanical probing can use acoustic signals~\cite{echoface,beyond,sonarguard,aface,ultraface}. These modalities respond to different physical properties: RF reflections are influenced by electromagnetic characteristics such as complex permittivity~\cite{permittivity}, whereas acoustic reflections are governed by mechanical characteristics such as acoustic impedance~\cite{acousticimpedance}. Such responses provide liveness cues beyond facial appearance and geometry.

Different from existing single-modal systems, \sysname combines electromagnetic and mechanical waves as complementary probing modalities to characterize the intrinsic physical responses of the presented medium. For the electromagnetic modality, we adopt mmWave sensing for its fine range--angle resolution and sensitivity to facial responses. Despite their complementary mechanisms, both modalities remain affected by the same user-dependent facial geometry.

\section{Preliminary Study}
\label{subsec:preliminary}
This section demonstrates the practical threat of high-fidelity 3D spoofing masks and provides empirical motivation for user-independent anti-spoofing with multimodal active probing.
\subsection{High-Fidelity 3D Mask Threat}
\label{sssec:maskattack}
\begin{table}[t]
\centering
\caption{Attack success against commercial and open-source face authentication systems. A checkmark denotes bypass; ASR is the success rate of high-fidelity 3D mask attacks.}
\label{tab:preliminary_camera}
\begin{tabular}{lccc}
\toprule
\multirow{2}{*}{\textbf{System}}
& \textbf{2D}
& \textbf{3D}
& \textbf{ASR}
\\
& \textbf{Attack}
& \textbf{Attack}
& \textbf{(\%)}
\\
\midrule

\multicolumn{4}{l}{\textit{COTS Devices}} \\
Smartphone Vendor A & $\times$     & $\checkmark$ & 63.33 \\
Smartphone Vendor B & $\times$     & $\checkmark$ & 83.33 \\
Smartphone Vendor C & $\times$     & $\checkmark$ & 36.67 \\
Tablet Vendor D     & $\times$     & $\checkmark$ & 73.33 \\

\midrule
\multicolumn{4}{l}{\textit{Open-Source Models}} \\
ArcFace~\cite{arcface}           & $\checkmark$ & $\checkmark$ & 95.61 \\
GhostFaceNet~\cite{ghostfacenet} & $\checkmark$ & $\checkmark$ & 92.34 \\
AdaFace ~\cite{adaFace}           & $\times$     & $\checkmark$ & 86.21 \\
FaceNet~\cite{facenet}           & $\times$     & $\checkmark$ & 56.32 \\

\bottomrule
\end{tabular}
\end{table}
We evaluate whether our high-fidelity 3D masks (see Sec.~\ref{sec:evaluation}) can bypass existing visual face authentication systems. For each system, a legitimate user first enrolls using their genuine face, after which another participant wears a high-fidelity 3D mask reproducing the enrolled user's appearance and facial geometry to launch the attack. An attack is considered successful when the attacker is accepted as the enrolled identity.

We first evaluate COTS face-unlock systems on three popular device models from each of four mainstream manufacturers, covering both smartphones and tablets. For ethical and responsible-disclosure considerations, we anonymize the manufacturers and exact device models. We conducted 30 attack attempts on each device. As shown in Table~\ref{tab:preliminary_camera}, the high-fidelity 3D masks successfully bypass commercial devices evaluated, achieving ASRs of 36.67\%–83.33\% (64.17\% on average), whereas conventional 2D attacks fail.
We further evaluate representative open-source face-recognition models. The high-fidelity 3D masks bypass all evaluated systems, whereas conventional 2D attacks fail against most of them. 

These results demonstrate that \textbf{the fabricated high-fidelity 3D masks reproduce sufficiently realistic facial appearance and geometry to pose a practical threat to commercial and open-source visual authentication systems.}
\begin{figure}[t]
    \centering

    \begin{subfigure}{0.326\linewidth}
        \centering
        \includegraphics[width=\linewidth]
        {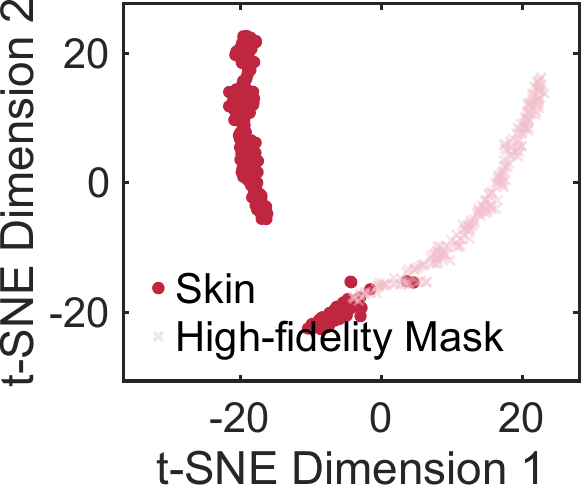}
        \caption{Acoustic.}
        \label{fig:controlled-acoustic}
    \end{subfigure}
    \hfill
    \begin{subfigure}{0.326\linewidth}
        \centering
        \includegraphics[width=\linewidth]
        {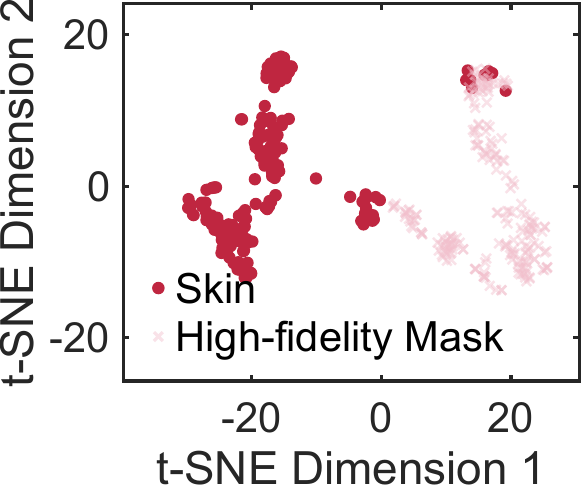}
        \caption{mmWave.}
        \label{fig:controlled-mmwave}
    \end{subfigure}
    \hfill
    \begin{subfigure}{0.326\linewidth}
        \centering
        \includegraphics[width=\linewidth]
        {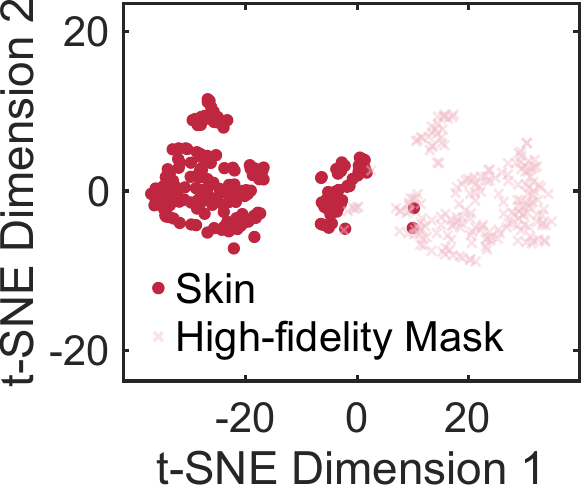}
        \caption{Fusion.}
        \label{fig:controlled-fusion}
    \end{subfigure}

    \caption{
        t-SNE visualizations under the same-user setting.
    }
    \label{fig:controlled-setting}
\end{figure}
\begin{figure}[t]
    \centering

    \begin{subfigure}{0.326\linewidth}
        \centering
        \includegraphics[width=\linewidth]
        {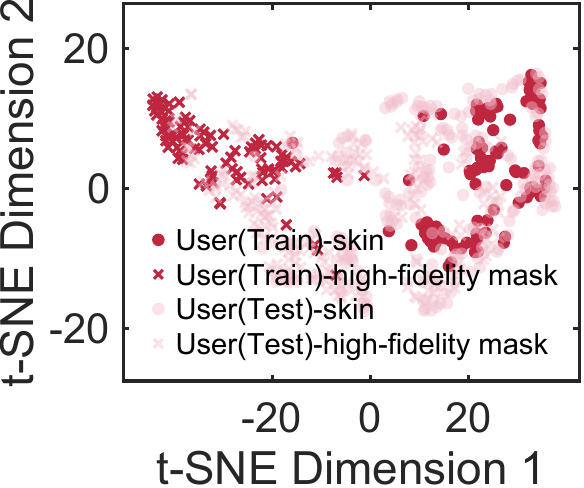}
        \caption{Acoustic.}
        \label{fig:cross-acoustic}
    \end{subfigure}
    \hfill
    \begin{subfigure}{0.326\linewidth}
        \centering
        \includegraphics[width=\linewidth]
        {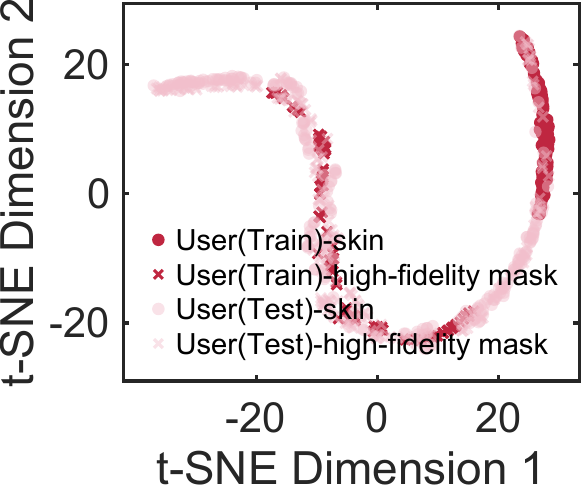}
        \caption{mmWave.}
        \label{fig:cross-mmwave}
    \end{subfigure}
    \hfill
    \begin{subfigure}{0.326\linewidth}
        \centering
        \includegraphics[width=\linewidth]
        {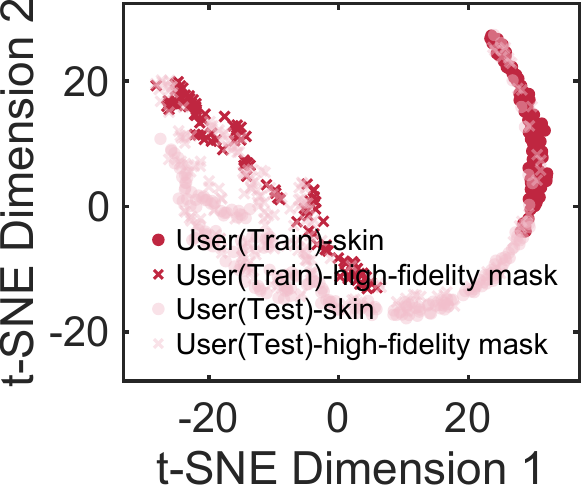}
        \caption{Fusion.}
        \label{fig:cross-fusion}
    \end{subfigure}

    \caption{
        t-SNE visualizations under the cross-user setting.
    }
    \label{fig:cross-user-setting}
\end{figure}
\begin{table}[t]
    \centering
    \caption{Classification performance of three methods.}
    \label{tab:preliminary_results}
    \begin{tabular}{lccc}
        \toprule
        \multirow{2}{*}{\textbf{Method}}
        & \textbf{Same-user}
        & \textbf{Cross-user}
        & \textbf{Drop} \\
        & \textbf{Acc. (\%)}
        & \textbf{Acc. (\%)}
        & \textbf{(\%)} \\
        \midrule
        Acoustic-only & 93.25 & 56.55 & 36.70$\downarrow$ \\
        mmWave-only   & 96.41 & 62.23 & 34.18$\downarrow$ \\
        Naive fusion  & 97.92 & 58.40 & 39.52$\downarrow$ \\
        \bottomrule
    \end{tabular}
\end{table}
\subsection{User-dependent Bias in Active Probing}
\label{sssec:opportunityandlimitation}
We then conduct preliminary experiments to examine the limitations of user-dependent active wireless probing. We evaluate three baseline configurations: mmWave-only sensing, acoustic-only sensing, and naive mmWave--acoustic fusion. For naive fusion, we apply attention-based fusion followed by a shared classification head~\cite{mmfas}. We synchronously collect 1,500 paired mmWave--acoustic samples from each of three participants under live-face and high-fidelity 3D mask conditions, following the acquisition setup described in Sec.~\ref{sec:evaluation}.

\noindent\textit{1) Physical sensitivity under same-user conditions.}
To verify that active probing can capture liveness-related physical responses, we evaluate skin and high-fidelity 3D mask samples from the same participant. For each configuration, we use ResNet18~\cite{resnet} as feature extractor, followed by MLP layers as classifier, and train them under the same protocol for fair comparison. As shown in Fig.~\ref{fig:controlled-setting}, both unimodal representations form distinct clusters for the two presentation media, and naive fusion also preserves this separation by combining complementary electromagnetic and mechanical responses. The results in Table~\ref{tab:preliminary_results} further show that all three baselines achieve high accuracy under the same-user setting. \textbf{These results confirm that both modalities capture liveness cues arising from the physical properties of the presented medium.}

\noindent\textit{2) Degradation across users.}
To examine whether user-dependent facial geometry limits the generalization of active probing, we evaluate the trained baselines on unseen participants while keeping the presented medium unchanged. As shown in Fig.~\ref{fig:cross-user-setting}, the feature distributions of all three baselines shift substantially across users, causing live and high-fidelity mask representations to increasingly overlap. Accordingly, the classification accuracy drops to around 50\%, as reported in Table~\ref{tab:preliminary_results}, which is close to random guessing for a binary task. This degradation arises because user-specific facial geometry alters the propagation paths and local reflection conditions observed by both modalities. \textbf{These results show that user-dependent variations remain deeply entangled with liveness cues, severely limiting cross-user generalization.}

\noindent\textit{3) Summary.}
These experiments reveal the limitations of active probing for user-independent face anti-spoofing. Although mmWave and acoustic measurements capture liveness-related physical responses, their representations remain strongly affected by user-dependent facial geometry, and multimodal fusion does not eliminate this interference. This motivates a user-independent representation that preserves liveness cues while suppressing user-dependent variations.

A straightforward way to improve cross-user generalization is to scale up training data. However, learning broadly generalizable representations, as in computer vision, would require data from a large and diverse population rather than more samples from the same users~\cite{imagenet}. More importantly, 3D face anti-spoofing would require fabricating a customized high-fidelity spoofing mask for each recruited user to provide corresponding live and spoof responses, with each customized mask costing approximately \$2,000. The resulting participant recruitment and user-specific mask fabrication introduce substantial cost and effort, making dataset scaling impractical.
\subsection{Shared Geometry across Modalities}
\label{subsec:shared_geometry}
\begin{figure}[t]
    \centering

    \begin{subfigure}{0.45\linewidth}
        \centering
        \includegraphics[width=\linewidth]
        {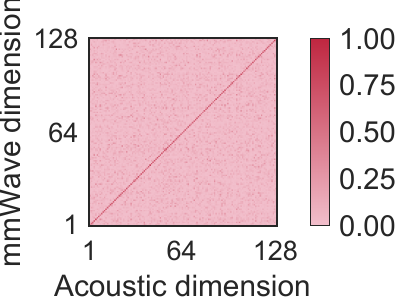}
        \caption{Same user condition.}
        \label{fig:cka-paired}
    \end{subfigure}
    \hfill
    \begin{subfigure}{0.45\linewidth}
        \centering
        \includegraphics[width=\linewidth]
        {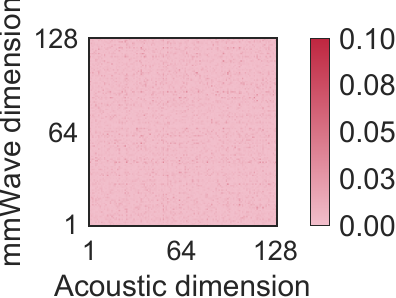}
        \caption{Cross user condition.}
        \label{fig:cka-unpaired}
    \end{subfigure}

    \caption{
        Cross-modal correlation matrices under same and cross
        user conditions.
    }
    \label{fig:representation_analysis}
\end{figure}
Since user-dependent facial geometry is preserved in bimodal representations and substantially degrades cross-user liveness detection, suppressing this shared nuisance information is essential for learning user-independent liveness cues.

\noindent\textit{Cross-modal consistency of user-dependent information.}
To validate our observation that the two co-located modalities encode consistent user-dependent variations, we conduct a cross-modal representation analysis. To make the semantic dimensions of the two modalities comparable, we first train the modality-specific encoders on paired mmWave and acoustic measurements, following the setup in Sec.~\ref{sssec:opportunityandlimitation}, using the Barlow Twins alignment objective~\cite{btloss}. To isolate user-dependent information, we then freeze the encoders and independently train lightweight identity-extracting and classification heads for each modality. Finally, to examine whether the extracted user-dependent representations share consistent geometric information across modalities, we compute their cross-modal similarity under same-user pairing and compare it with that under randomly paired different-user samples. Random pairing preserves the marginal feature distributions while disrupting user correspondence. As shown in Fig.~\ref{fig:representation_analysis}, same-user pairs exhibit strong cross-modal similarity, whereas the similarity approaches zero under different-user pairing. \textbf{These results demonstrate that the aligned mmWave and acoustic representations encode consistent variations induced by the same user-dependent facial geometry.}

\sssec{$\blacksquare$\textbf{Key insight:}} 
The \textit{co-located} sensors probe the face from nearly the same viewpoint and therefore exhibit shared user-dependent responses dominated by facial geometry.

\subsection{Motivation}
\label{subsec:general_motivation}

The above observation motivates a different way of leveraging multimodal sensing.
Conventional multimodal fusion aggregates information from different modalities, preserving not only complementary liveness cues but also the user-dependent variations shared across them. In contrast, \sysname exploits this shared information as a nuisance factor that can be suppressed through cross-modal comparison.

We describe the received facial response using the following task-level coupling model:
\begin{equation}
R^p =
\mathcal{H}^p
\left(
S, \eta^p
\right)
+
\xi^p,
\quad
p \in \{\mathrm{mm},\mathrm{ac}\},
\label{eq:facial_response}
\end{equation}
where $S$ represents user-specific facial geometry, $\eta^p$ denotes the modality-specific physical response associated with the presented medium, and $\xi^p$ represents environmental multipath and measurement noise. The function $\mathcal{H}^p(\cdot)$ denotes the modality-specific sensing process that couples these factors into the observed response. The objective of \sysname is to extract liveness-discriminative information induced by $\eta^p$ while suppressing nuisance variations associated with $S$. 
Since the co-located modalities experience the same facial geometry but respond to different physical properties of the presented medium, $S$ tends to induce cross-modal shared variations, whereas $\eta^p$ gives rise to modality-specific responses. 
\textbf{This motivates suppressing cross-modal shared user-dependent geometry information while preserving modality-specific liveness cues for user-independent face anti-spoofing.}

\section{Threat Model}
\label{subsec:threat_model}

We consider a physical presentation attacker who attempts to impersonate a legitimate user using a high-fidelity 3D spoofing mask. The attacker presents the mask through the normal face-authentication interface. In an end-to-end face-authentication system, a successful impersonation requires the spoofing mask to both pass liveness detection and be matched to the target user by the face-recognition module. As the liveness detection component, \sysname focuses on determining whether the presented face is live or spoofed. We make the following assumptions regarding the attacker’s capabilities and the deployment setting.
 
\noindent\textit{1) Spoof fabrication capabilities.}
The attacker may obtain multi-view images from social media or other public sources and reconstruct the target's 3D facial geometry using commercially available tools. A high-fidelity mask can then be fabricated through printing, molding, or casting using diverse materials, such as resin, latex, silicone, and other polymers.

\noindent\textit{2) Real world dployment.}
We do not assume that the sensing signals interact exclusively with facial skin or the spoofing material. The sensed facial region may encompass other materials such as hair, eyeglasses, or face coverings. \sysname must therefore distinguish live and spoof presentations despite interference from multiple materials.

\noindent\textit{3) Knowledge and system access.}
The attacker can physically present a high-fidelity 3D mask to the authentication system and observe the final authentication outcome. However, the attacker cannot access or modify internal sensing measurements, model parameters, intermediate representations, hardware, software, or inject forged sensing signals. 

\noindent\textit{4) Assumption of the visual pipeline.}
\sysname operates alongside a camera-based authentication module. The visual pipeline can simply reject obvious attacks with visible mask boundaries or inconsistent uncovered regions and triggers \sysname when a face is detected within the predefined region of interest (ROI). \sysname then performs liveness detection solely using actively probed mmWave and acoustic responses, without relying on visual information. It can therefore serve as a plug-in liveness detection module independent of camera configuration and illumination conditions.

\section{Methodology}
\label{sec:method}
\begin{figure*}[t]
    \centering
    \includegraphics[width=\linewidth]{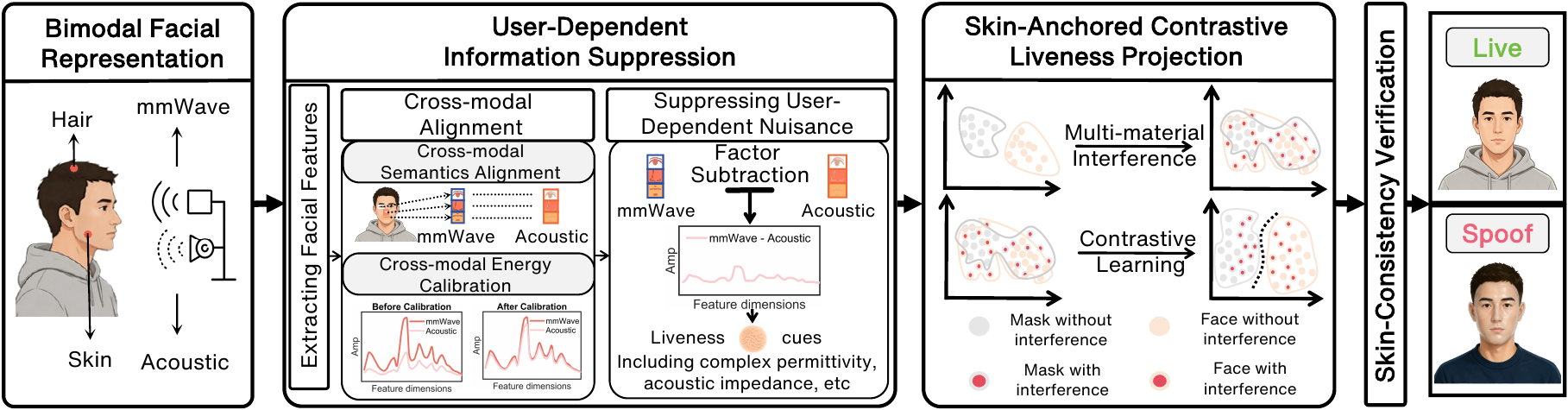}
    \caption{Overview of \sysname, which comprises three main components. Bimodal facial representation constuction; user-dependent information suppression removes shared geometry to obtain liveness representations; and skin-anchored contrastive projection recovers reliable liveness information under material variations for final live/spoof classification.}
    \label{fig:overview}
\end{figure*}
In this section, we present \sysname, a multimodal framework for user-independent 3D face anti-spoofing, as illustrated in Fig.~\ref{fig:overview}. Given synchronized mmWave and acoustic reflections from a presented face, \sysname determines whether the presentation is live or spoofed without requiring user-specific signal enrollment. The framework consists of three stages. First, \sysname constructs multipath-suppressed mmWave range--azimuth and 2D-AoA responses, together with an acoustic channel impulse response (Sec.~\ref{sec:method-facial-response}). Second, it aligns and calibrates the heterogeneous representations, suppresses their shared user-dependent variations (Sec.~\ref{sec:method-suppression}). Third, under material variations, \sysname uses skin-anchored contrastive projection (Sec.~\ref{sec:method-multimaterial}) and skin-consistency verification for robust live/spoof classification (Sec.~\ref{subsec:classification}).

\subsection{Bimodal Facial Representation}
\label{sec:method-facial-response}

Given synchronized mmWave signals and acoustic recordings reflected from a presented face, \sysname converts the raw measurements into modality-specific facial response representations while reducing environmental multipath before subsequent cross-modal processing.

\noindent\textit{1) mmWave facial response.}
\sysname employs an FMCW mmWave radar to measure electromagnetic reflections from the presented face. After range processing and array-based angle estimation, it constructs a range--azimuth and a 2D angle-of-arrival (2D-AoA) map. These representations characterize how facial reflections are distributed. They are jointly affected by the presented surface, user-dependent facial geometry $S$.

\noindent\textit{2) Acoustic facial response.}
For acoustic sensing, \sysname transmits broadband Zadoff--Chu (ZC) sequences and correlates the received echoes with the transmitted signal to estimate the channel impulse response (CIR). Owing to the autocorrelation property of ZC sequences~\cite{zcgood}, the CIR resolves reflections from different propagation paths around the presented face. 

\noindent\textit{3) Multipath reduction.}
The constructed representations also contain weak reflections and environmental multipath from surrounding objects. \sysname applies modality-specific filtering to suppress such interference. For mmWave sensing, it identifies the maximum-energy peak in each facial response representation and sets a relative energy threshold with respect to this peak. Components below the threshold are removed, retaining the dominant facial reflections in the range--azimuth and 2D-AoA maps. For acoustic sensing, \sysname locates the strongest peak in the CIR and uses its time of flight as the reference. It retains CIR components within a local ToF window for this peak and suppresses reflections outside the window.

\subsection{User-Dependent Information Suppression}
\label{sec:method-suppression}
Environmental multipath suppression removes reflections from surrounding objects, but the retained facial responses are still jointly affected by the user-dependent facial geometry $S$. In user-independent face anti-spoofing, the latter factors are nuisance variables: a classifier may associate particular facial structure with the live/spoof label instead of learning generalizable liveness cues.

\sysname suppresses these nuisance factors by exploiting a key property established in Sec.~\ref{subsec:general_motivation}: co-located mmWave and acoustic sensors encode correlated user-dependent variations, while their liveness responses arise from different physical mechanisms. Therefore, the suppression process consists of feature extraction, cross-modal alignment and cross-modal subtraction for user-dependent nuisance
factor.

\paragraph{1) Extracting facial features.}
Let $\mathbf{x}^{p}$ denote the multipath-suppressed input of modality $p$, where $\mathbf{x}^{\mathrm{mm}}$ collectively represents the range--azimuth and 2D-AoA maps of mmWave, and $\mathbf{x}^{\mathrm{ac}}$ represents the acoustic CIR. Modality-specific ResNet-based encoders~\cite{1dres,resnet} transform these inputs into high-level representations $\bar{\mathbf{k}}^{\mathrm{mm}}$ and $\bar{\mathbf{k}}^{\mathrm{ac}}$, respectively. Then the task-relevant information encoded in each representation can be defined as:
\begin{equation}
\bar{\boldsymbol{k}}^{p}
\approx
\underbrace{
\boldsymbol{g}^{p}(S)
}_{\substack{\text{Geometry}}}
+
\underbrace{
\boldsymbol{\ell}^{p}\!\left(\eta^{p}\right)
}_{\substack{\text{Liveness cues}}},
\quad
p \in \{\mathrm{mm},\mathrm{ac}\},
\label{eq:general_idea}
\end{equation}
where $\boldsymbol{g}^{p}(S)$ denotes the modality-specific representation induced by user-dependent facial geometry $S$, while $\boldsymbol{\ell}^{p}(\eta^{p})$ denotes the modality-specific liveness information induced by the physical response $\eta^{p}$. Although the two modalities encode these factors differently, both are influenced by the same underlying facial geometry. 
\textit{The additive form in Eq.~\ref{eq:general_idea} is not intended as a strictly linear physical model, but rather as a target representation structure that our subsequent learning objectives encourage the feature extractors to approximate.}

\paragraph{2) Cross-modal alignment.}
After modality-specific feature extraction, the mmWave and acoustic representations cannot be directly compared because of two modality-induced mismatches. First, their latent dimensions are not semantically aligned: a dimension encoding a particular facial variation in one modality may correspond to a different factor in the other. Second, their feature magnitudes may differ substantially because of heterogeneous propagation losses, hardware responses, and encoder activation distributions. Direct subtraction would therefore mix semantic and scale mismatches with the desired cross-modal differences.

To address the semantic mismatch, \sysname first employs modality-specific alignment heads to map $\bar{\mathbf{k}}^{\mathrm{mm}}$ and $\bar{\mathbf{k}}^{\mathrm{ac}}$ into a common semantic space, yielding $\tilde{\mathbf{k}}^{\mathrm{mm}}$ and $\tilde{\mathbf{k}}^{\mathrm{ac}}$. The alignment heads are jointly optimized using $\mathcal{L}_{\mathrm{align}}$ (Barlow Twins objective~\cite{btloss}), which encourages correlated cross-modal factors to occupy corresponding latent dimensions.

\begin{figure}[t]
    \centering

    \begin{subfigure}[t]{0.506\linewidth}
        \centering
        \includegraphics[width=\linewidth]{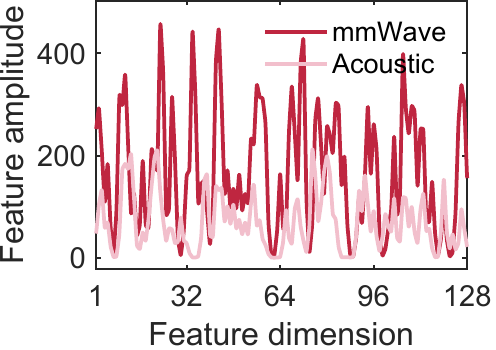}
        \caption{Before calibration.}
        \label{fig:beforecalibration}
    \end{subfigure}
    \hfill
    \begin{subfigure}[t]{0.47\linewidth}
        \centering
        \includegraphics[width=\linewidth]{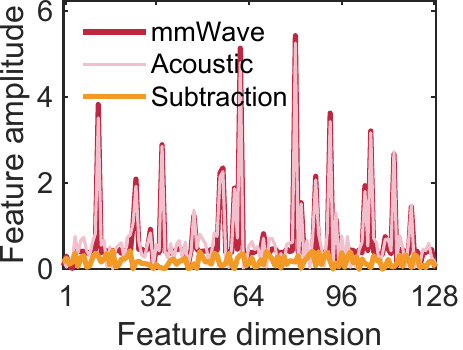}
        \caption{After calibration.}
        \label{fig:aftercalibration}
    \end{subfigure}

    \caption{Cross-modal features before and after calibration. Semantic alignment yields correlated but scale-mismatched features, while energy  calibration makes them comparable.}
    \label{fig:calibration}
\end{figure}

Semantic alignment does not ensure comparable feature magnitudes as shown in Fig.~\ref{fig:beforecalibration}. \sysname therefore further applies cross-modal attention~\cite{crossattention} based head to perform dimension-wise energy calibration, producing calibrated representations $\hat{\mathbf{k}}^{\mathrm{mm}}$ and $\hat{\mathbf{k}}^{\mathrm{ac}}$. This step reduces systematic scale imbalance while preserving informative modality-specific differences.

After cross-modal alignment, as shown in Fig.~\ref{fig:aftercalibration}, the geometry-related components of the two modalities become approximately comparable in both latent coordinates and feature scales. The network architecture is illustrated in Fig.~\ref{fig:method}.
\paragraph{3) Cross-modal subtraction for user-dependent nuisance factor.}
As shown in Fig.~\ref{fig:aftercalibration}, semantic alignment and energy calibration make the geometry-related components of the two modalities approximately comparable. However, these dominant components can obscure the weaker modality-specific differences associated with liveness. In the figure, the high-amplitude dark- and light-red curves represent the similar bimodal features, while their lower-amplitude residual, shown in orange, preserves modality-specific liveness cues. To suppress the dominant shared geometry, \sysname therefore applies cross-modal subtraction. Based on Eqs.~\ref{eq:general_idea}, the main liveness branch is constructed as
\begin{equation}
\mathbf{k}^{L}
=
\Psi_{L}\!\left(
\hat{\mathbf{k}}^{\mathrm{mm}}
-
\hat{\mathbf{k}}^{\mathrm{ac}}
\right),
\label{eq:liveness_disentanglement}
\end{equation}
where $\Psi_L$ denotes the liveness head and $\mathbf{k}^{L}$ captures the residual modality-specific liveness information after suppressing shared geometry-related responses. In parallel, we apply an auxiliary geometry head $\Psi_G$ to the sum of the calibrated bimodal features, producing $\mathbf{k}^{G}$ that emphasizes information consistently expressed across modalities and serves as a proxy for user-dependent geometry.

To further disentangle liveness and geometry representations while preventing information loss, \sysname imposes two constraints and defines the subtraction objective as
\begin{equation}
\begin{aligned}
\mathcal{L}_{\mathrm{sub}}
={}&
\operatorname{sim}^{2}(\mathbf{k}^{L},\mathbf{k}^{G}) \\
&+
\operatorname{MSE}\!\left(
\Psi_{\mathrm{res}}(\mathbf{k}^{L}+\mathbf{k}^{G}),
\hat{\mathbf{k}}^{\mathrm{mm}}+\hat{\mathbf{k}}^{\mathrm{ac}}
\right).
\end{aligned}
\label{eq:sub_loss}
\end{equation}
where $\operatorname{sim}(\cdot,\cdot)$ denotes cosine similarity, $\operatorname{MSE}(\cdot,\cdot)$ is the mean squared error, and $\Psi_{\mathrm{res}}(\cdot)$ denotes a lightweight reconstruction head. The first term imposes a decorrelation constraint between $\mathbf{k}^{L}$ and $\mathbf{k}^{G}$, reducing their information overlap and limiting user-dependent information retained in the liveness branch. The second term imposes a reconstruction constraint, requiring the two representations to remain jointly informative about the calibrated multimodal representation and preventing degenerate solutions. Together, these constraints encourage cross-modal subtraction to suppress shared user-dependent variations while preserving liveness-discriminative information.

\subsection{Skin-Anchored Contrastive Liveness Projection}
\label{sec:method-multimaterial}

\begin{figure}[t]
    \centering

    \begin{subfigure}[t]{0.495\linewidth}
        \centering
        \includegraphics[width=\linewidth]{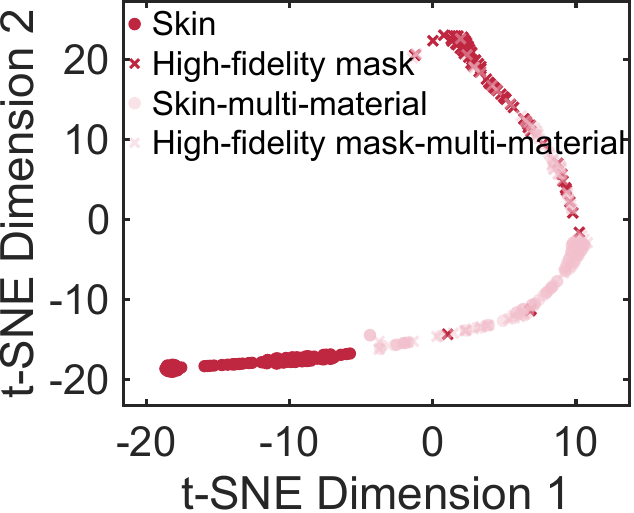}
        \caption{Material-induced feature drift.}
        \label{fig:tsne_stage1}
    \end{subfigure}
    \hfill
    \begin{subfigure}[t]{0.495\linewidth}
        \centering
        \includegraphics[width=\linewidth]{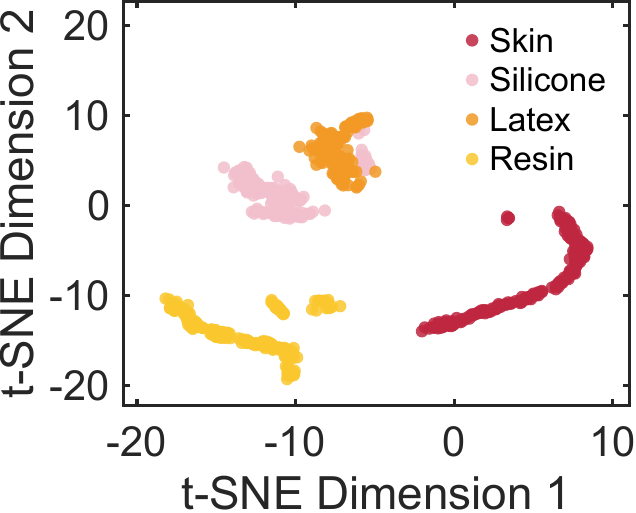}
        \caption{Stable skin distribution.}
        \label{fig:gaussian}
    \end{subfigure}

    \caption{Multi-material interference causes feature drift, while genuine-skin representations remain stable across users.}
    \label{fig:tsne_stage}
\end{figure}

In this section, \sysname addresses two practical issues: multi-material induced drift in the liveness representation $\mathbf{k}^{L}$ and the open-ended diversity of spoofing materials. 
Additional materials introduce modality-specific physical responses into $\mathbf{k}^{L}$, causing representation drift, whereas the diversity of spoofing materials precludes exhaustive material-specific modeling.
Therefore, we first characterize how material composition biases $\mathbf{k}^{L}$, then identify stable genuine-skin features, and finally introduce skin-anchored contrastive learning to project liveness-discriminative representations. 

\paragraph{1) Feature drift caused by material composition.}
To characterize how additional materials bias the liveness representation $\mathbf{k}^{L}$, we examine material variations under a fixed-user setting. We first train the model using clean genuine-skin and silicone-mask samples from the same user. Since facial measurements are inevitably affected by hair and eyebrows, we use the palm to obtain an uncontaminated genuine-skin response in this controlled analysis. We then evaluate the user with face coverings, eyeglasses, and hairstyle changes. As shown in Fig.~\ref{fig:tsne_stage1}, these variations shift both genuine and spoofing representations away from their clean counterparts. This occurs because different materials exhibit distinct electromagnetic and mechanical responses that, unlike shared geometric variations, are modality-specific and can remain after cross-modal subtraction. Such feature drift may therefore lead to incorrect live/spoof decisions.

\paragraph{2) Stable skin features.}
To understand whether material-dependent physical responses exhibit consistent patterns in the liveness representation $\mathbf{k}^{L}$, we analyze the feature distributions across different users and spoofing materials. We use the model as in the previous experiment, and then visualize $\mathbf{k}^{L}$ from additional participants with different ages, genders, and racial backgrounds, together with spoofing masks made of silicone, latex, and resin. As shown in Fig.~\ref{fig:gaussian}, we obtain three observations: (1) samples made of the same material naturally form compact clusters in the feature space; (2) genuine-skin representations from different users show highly consistent and same distributions despite demographic variations; and (3) materials with similar physical properties may exhibit partially overlapping distributions, while remaining separated from the genuine-skin distribution. These observations indicate that genuine skin provides a stable reference, whereas the spoof-material space is heterogeneous. \sysname therefore focuses on modeling genuine-skin characteristics instead of exhaustively characterizing individual spoofing materials.

\paragraph{3) Contrastive-based liveness projection.}
Building on the above observations, to mitigate material-induced feature drift and avoid exhaustively modeling the open-ended spoof-material space, \sysname introduces a skin-anchored contrastive projection. This design is motivated by two considerations. First, a projection head optimized with contrastive supervision encourages the projected space to discard variations irrelevant to the contrastive target~\cite{projection}, thereby suppressing responses introduced by face coverings, eyeglasses, hair, and other materials. Second, conventional pairwise contrastive learning only constrains local relationships between sampled pairs, while inheriting sample-specific variations or outliers. We therefore introduce a skin anchor $\mathbf{c}_{\mathrm{skin}}$, constructed from clean genuine skin, as a population-level prototype of stable genuine-skin characteristics~\cite{anchor}. Consequently, \sysname preserves skin-related liveness evidence under material interference without modeling individual spoofing materials.

Specifically, a shared projection head $\Phi_{\mathrm{pro}}(\cdot)$ maps the geometry-suppressed liveness representation $\mathbf{k}^{L}$ into the projected representation
$\mathbf{k}^{L'}=\Phi_{\mathrm{pro}}(\mathbf{k}^{L})$.
The skin anchor $\mathbf{c}_{\mathrm{skin}}$ is constructed from clean genuine-skin representations in the same projected space, while its training-time update is described in Sec.~\ref{subsec:classification}. Let $\mathcal{R}$ and $\mathcal{F}$ denote the genuine and spoofing samples in a training batch, respectively. We define
$q(\mathbf{u},\mathbf{v})
=
\exp(\operatorname{sim}(\mathbf{u},\mathbf{v})/\tau)$,
where $\operatorname{sim}(\cdot,\cdot)$ denotes cosine similarity and $\tau$ is the temperature parameter. The skin-anchored contrastive objective is
\begin{equation}
\mathcal{L}_{\mathrm{con}}
=
-\frac{1}{|\mathcal{R}|}
\sum_{i\in\mathcal{R}}
\log
\frac{
q(\mathbf{k}^{L'}_i,\mathbf{c}_{\mathrm{skin}})
}{
q(\mathbf{k}^{L'}_i,\mathbf{c}_{\mathrm{skin}})
+
\sum_{j\in\mathcal{F}}
q(\mathbf{k}^{L'}_i,\mathbf{k}^{L'}_j)
}.
\label{eq:multimaterial_contrastive}
\end{equation}

For each genuine presentation, $\mathbf{c}{\mathrm{skin}}$ serves as the positive reference, while spoofing representations in the batch serve only as negatives. This objective pulls material-affected genuine samples toward the clean-skin anchor, suppressing feature shifts introduced by additional materials. Consequently, $\Phi{\mathrm{pro}}(\cdot)$ learns to preserve genuine-skin liveness evidence while suppressing material-induced variations, without explicitly modeling the diverse spoof-material space.

\subsection{Loss Construction and Training}
\label{subsec:classification}

This section defines \sysname's training objective and strategy. Because spoofing materials form a diverse evolving space, the classification objective is formulated around consistency with skin rather than material-specific recognition. The overall framework is optimized in two stages: Stage I learns the initial clean user-independent liveness representation, while Stage II adapts the projected space to material variations.

\paragraph{1) Skin-consistency classification objective.}
Because spoofing materials are open-ended and cannot be exhaustively covered during training, a conventional binary classifier may learn a decision boundary tied to the observed spoofing distributions. \sysname therefore formulates liveness classification as skin-consistency estimation: whether the projected representation $\mathbf{k}^{L'}$ preserves genuine-skin characteristics relative to the skin anchor $\mathbf{c}_{\mathrm{skin}}$. We quantify this consistency as:
\begin{equation}
s_{\mathrm{skin}}
=
\operatorname{sim}
\left(
\mathbf{k}^{L'},
\mathbf{c}_{\mathrm{skin}}
\right).
\label{eq:skin_similarity}
\end{equation}
We use cosine similarity, $\operatorname{sim}(\cdot,\cdot)$, rather than Euclidean distance because genuine-skin features may vary in magnitude across users while exhibiting similar directions. Cosine similarity is therefore less sensitive to magnitude variations and better reflects the shared skin characteristics across users.

To distinguish whether the skin-consistency score is sufficiently high for acceptance, a decision threshold is required. Instead of manually selecting a fixed threshold, which may be sensitive to the learned score distribution and generalize poorly across conditions, \sysname learns the acceptance boundary from data using a lightweight calibration function:
\begin{equation}
P(\mathrm{live})
=
\sigma\!\left(
\alpha(s_{\mathrm{skin}}-\delta)
\right),
\label{eq:skin_calibration}
\end{equation}
where $\delta$ denotes the learnable skin-consistency boundary, $\alpha>0$ controls the transition sharpness, and $\sigma(\cdot)$ is the sigmoid function. We then define the classification objective as
\begin{equation}
\mathcal{L}_{\mathrm{cla}}
=
\operatorname{BCE}
\left(
P(\mathrm{live}),y
\right),
\label{eq:calibration_loss}
\end{equation}
where $\operatorname{BCE}(\cdot,\cdot)$ denotes the binary cross-entropy loss and $y$ is the ground-truth liveness label. 

Unlike an unrestricted classification head over $\mathbf{k}^{L'}$, $\mathcal{L}_{\mathrm{cla}}$ constrains prediction through similarity to the genuine-skin reference. It therefore learns how much genuine-skin consistency is required for acceptance. At inference time, the same calibrated score is used directly, with $P(\mathrm{live})\geq0.5$ classified as live and otherwise as spoof.
\begin{figure}
    \centering
    \includegraphics[width=\linewidth]{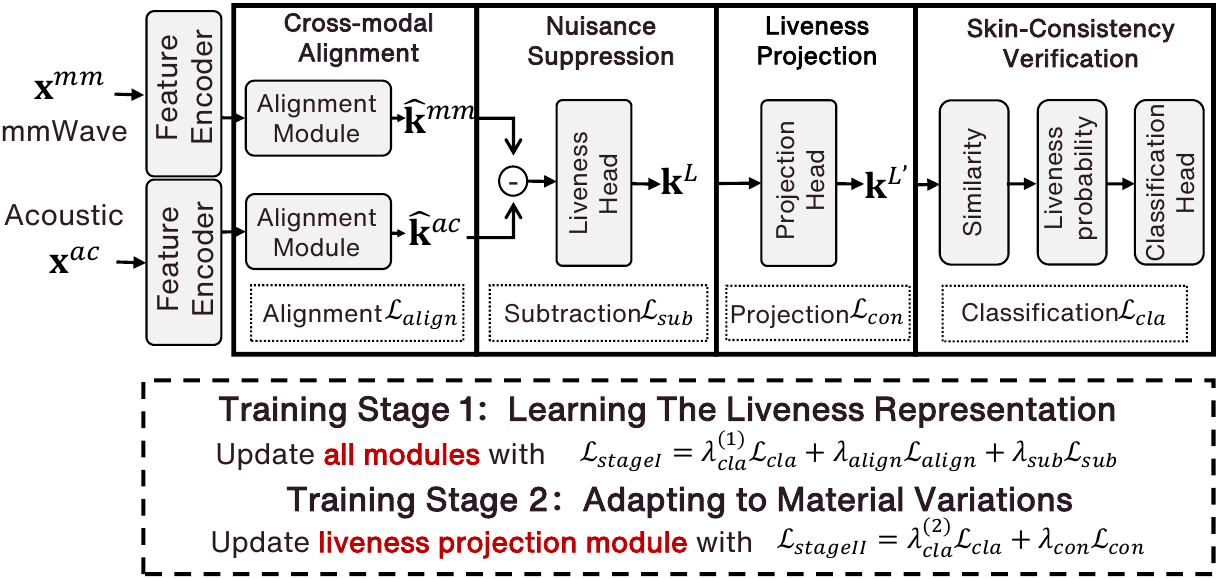}
    \caption{Net freamwork and training process.}
    \label{fig:method}
\end{figure}

\paragraph{2) Training.}
To optimize \sysname, we adopt a two-stage training strategy, as shown in Fig.~\ref{fig:method}. Stage I learns the core user-independent liveness representation by jointly optimizing the feature encoder, cross-modal alignment, subtraction, and projection modules. Stage II refines the projected representation using skin-anchored objectives to improve robustness to material variations. We next describe the two stages in detail.

\noindent\textit{Stage I: Learning the liveness representation.}
To establish an initial genuine-skin space, Stage I trains the network on clean single-material presentations. The skin anchor $\mathbf{c}_{\mathrm{skin}}$ is jointly optimized as a learnable prototype of genuine-skin characteristics. The objective is
\begin{equation}
\mathcal{L}_{\mathrm{stageI}}
=
\lambda_{\mathrm{cla}}^{(1)}\mathcal{L}_{\mathrm{cla}}
+
\lambda_{\mathrm{align}}\mathcal{L}_{\mathrm{align}}
+
\lambda_{\mathrm{sub}}\mathcal{L}_{\mathrm{sub}}.
\label{eq:base_loss}
\end{equation}
where $\lambda$ is the coefficient weight. Stage I therefore learns a clean geometry-suppressed skin representation.

\noindent\textit{Stage II: Adapting to material variations.}
To learn liveness representations robust to material variations, Stage II freezes all modules except the projection head, which is updated using multi-material presentations. Since updating $\Phi_{\mathrm{pro}}(\cdot)$ changes the distribution of liveness representations $\mathbf{k}^{L'}$, the Stage-I skin anchor may become inconsistent with the evolving projected space. We therefore update $\mathbf{c}_{\mathrm{skin}}$ using an exponential moving average (EMA) over clean genuine-skin representations:
$\mathbf{c}_{\mathrm{skin}}^{(t)}
=
m\mathbf{c}_{\mathrm{skin}}^{(t-1)}
+
\frac{1-m}{|\mathcal{C}_{t}|}
\sum_{i\in\mathcal{C}_{t}}
\mathbf{k}^{L'}_{i}$,
where $\mathcal{C}_{t}$ denotes the clean genuine-skin samples in batch $t$ and $m$ is the momentum coefficient. The Stage-II objective is
\begin{equation}
\mathcal{L}_{\mathrm{stageII}}
=
\lambda_{\mathrm{cla}}^{(2)}\mathcal{L}_{\mathrm{cla}}
+
\lambda_{\mathrm{con}}\mathcal{L}_{\mathrm{con}}.
\label{eq:multimaterial_loss}
\end{equation}
Through this optimization, the projection head learns to recover clean-skin-related liveness characteristics from material-affected representations. As a result, the projected representation becomes robust to practical material variations without explicitly modeling individual spoofing materials.

Overall, the training strategy yields a liveness representation for reliable user-independent facial anti-spoofing.
\section{Evaluation}
\label{sec:evaluation}
\subsection{Experimental Setup}
\label{subsec:experimental_setup}
\begin{figure}[t]
    \centering

    \begin{subfigure}[t]{0.502\linewidth}
        \centering
        \includegraphics[width=\linewidth]{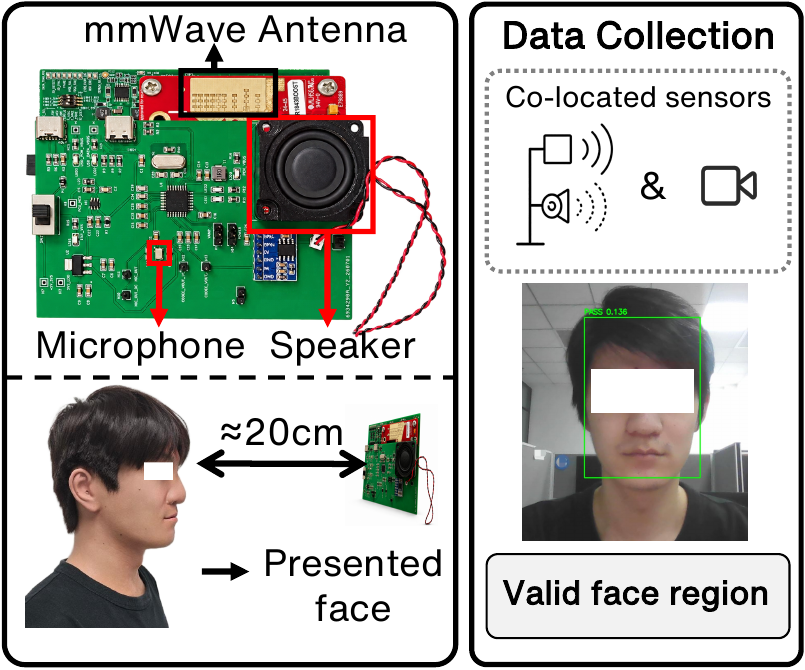}
        \caption{Prototype and data collection.}
        \label{fig:device}
    \end{subfigure}
    \hfill
    \begin{subfigure}[t]{0.48\linewidth}
        \centering
        \includegraphics[width=\linewidth]{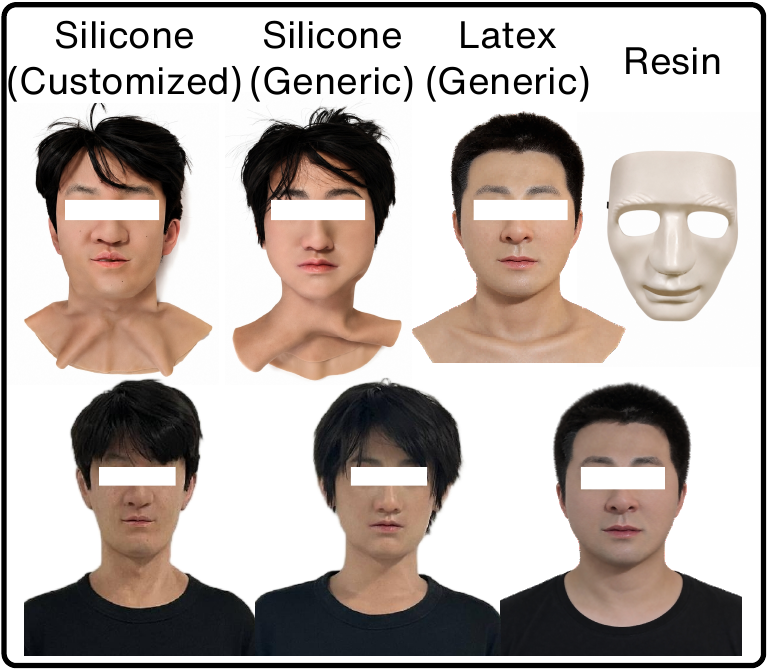}
        \caption{Experiment mask.}
        \label{fig:mask}
    \end{subfigure}

    \caption{Experimental setup of \sysname.}
    \label{fig:setup}
\end{figure}

We implement synchronized multimodal data acquisition and signal processing in MATLAB R2024b, extracting range--azimuth heatmaps and 2D-AoA spectra from mmWave signals and CIRs from acoustic signals. Model training and inference are performed on a server equipped with an NVIDIA RTX 4090 GPU with 24~GB memory.

\noindent\textbf{1) System implementation.}
As shown in Fig.~\ref{fig:device}, \sysname uses a TI IWR1843BOOST mmWave radar~\cite{TI_IWR1843,TI_DCA1000EVM} operating at 76--81~GHz and a co-located speaker--microphone pair transmitting acoustic signals at 17--22~kHz. The two modalities synchronously probe the presented face and capture complementary physical responses, respectively.

\noindent\textbf{2) High-fidelity spoofing masks.}
To construct realistic 3D spoofing attacks, we fabricate customized masks from 3D facial models, as shown in Fig.~\ref{fig:mask}. The spoofing masks are made of silicone, latex, and resin. The silicone masks include 
one customized mask reproducing the facial appearance of one of our authors who is not among the 14 participants and one generic-mask based on the manufacturer's template facial model. 
The generic silicone mask provides a 3D spoofing reference for learning the physical differences between genuine skin and spoofing materials, while the customized silicone mask simulates high-fidelity impersonation attacks to ensure that both the mask wearer and the impersonated identity are unseen during training. Latex and resin masks introduce new spoofing materials to evaluate generalization.

\noindent\textbf{3) Data collection.}
We recruit 14 participants aged 18–45 years, covering both male and female participants with diverse demographic backgrounds and skin characteristics.
We follow the data-collection procedure in Fig.~\ref{fig:device}. Each participant faces \sysname at a distance about 20~cm. A Haar Cascade detector~\cite{haar} detects the face, and synchronized mmWave--acoustic acquisition is triggered when at least 85\% of the detected face lies within the predefined region of interest (ROI). Each trial then collects a synchronized multi-modal frame.

 For each participant, we first collect 500 trials of clean palm skin and 500 generic silicone-mask trials for Stage I training, which are used to train the main network and learn the clean-skin representation space.
For Stage II, we collect genuine-face and generic-mask presentations under three facial conditions: no accessory, face covering, and eyeglasses, with 250 trials for each presentation–condition combination. These samples are used to train the projection head to identify genuine-skin characteristics in the presence of additional facial materials.
Finally, we collect 250 trials of the customized mask with no facial accessory, which are reserved for unseen-user testing. Overall, each participant contributes 2,750 bimodal trials, resulting in 38,500 trials across 14 participants.

\noindent\textbf{4) Evaluation protocol.}
We evaluate \sysname under two main scenarios.
(1) \textit{Registered-user scenario.} This setting evaluates performance when all user identities are observed during training. We randomly split the data of all 14 participants into 80\% for training and 20\% for testing, such that every participant appears in both sets. Results are averaged over seven random splits.
(2) \textit{Unseen-user scenario.} This setting evaluates cross-user generalization using seven-fold cross-validation. In each fold, 12 participants are used exclusively for training and the remaining two exclusively for testing. 
Both training stages use only samples from the 12 training participants. Testing is performed on genuine-face and customized silicone-mask presentations from the two held-out participants. Therefore, for spoofing trials, both the mask wearer and the identity impersonated by the customized mask are strictly unseen during training. Moreover, all impact-factor experiments are conducted using the models trained in unseen user scenario, and results are averaged across the seven folds.

\noindent\textbf{5) Baselines and metrics.}
We compare \sysname with vision-only, acoustic-only, mmWave-only, and multimodal baselines. \textit{Vision-only} uses MiniFAS~\cite{silentface}, while \textit{Acoustic-only} implements AFace~\cite{aface}. Since existing mmWave-based face anti-spoofing methods do not directly address user-independent detection, we adopt MID~\cite{mid}, a representative mmWave material recognition method, as the \textit{mmWave-only} baseline. For the multimodal baseline, we adapt mmFAS~\cite{mmfas} by replacing its visual inputs with mmWave and acoustic representations while retaining its fusion architecture. We report accuracy (ACC), false acceptance rate (FAR), false rejection rate (FRR), equal error rate (EER), area under the ROC curve (AUC), and F1 score. ACC measures overall classification accuracy; FAR and FRR quantify spoof acceptance and genuine rejection errors, respectively; EER is the operating point where FAR equals FRR; AUC summarizes discrimination across thresholds; and F1 balances precision and recall.

\subsection{Overall Performance}
\label{subsec:overall_performance}
This section evaluates \sysname under registered and unseen user settings. We first assess liveness classification and cross-user generalization, then validate cross-modal subtraction for suppressing user-dependent information, and finally quantify the effectiveness of each module through ablation studies.

\begin{table*}[t]
\centering
\caption{Overall comparison and ablation results under the
registered and unseen user settings.
$\uparrow$ indicates higher is better, and $\downarrow$ indicates
lower is better. ACC, FAR, FRR, EER, and F1 are reported in
percentage (\%), while AUC is reported in the range $[0,1]$.}
\label{tab:overall_performance}

\begin{tabular*}{\textwidth}{
@{\extracolsep{\fill}}lcccccc|cccccc@{}
}
\toprule
\multirow{2}{*}{Method}
& \multicolumn{6}{c|}{Registered User}
& \multicolumn{6}{c}{Unseen User} \\
\cmidrule(lr){2-7}
\cmidrule(lr){8-13}

& ACC$\uparrow$
& FAR$\downarrow$
& FRR$\downarrow$
& EER$\downarrow$
& AUC$\uparrow$
& F1$\uparrow$
& ACC$\uparrow$
& FAR$\downarrow$
& FRR$\downarrow$
& EER$\downarrow$
& AUC$\uparrow$
& F1$\uparrow$ \\

\midrule

MiniFAS~\cite{silentface}
& 52.00 & 95.00 & 1.00 & 48.97 & 0.52 & 67.35
& 50.40 & 99.00 & 0.20 & 49.80 & 0.50 & 66.80 \\

AFace~\cite{aface}
& 96.50 & 2.42 & 4.58 & 3.75 & 0.99 & 96.46
& 50.50 & 44.40 & 54.60 & 51.10 & 0.54 & 47.84 \\

MID~\cite{mid}
& 94.00 & 3.50 & 8.50 & 6.67 & 0.97 & 93.85
& 56.10 & 16.20 & 71.60 & 42.60 & 0.61 & 39.28 \\

mmFAS~\cite{mmfas}
& 98.79 & 0.83 & 1.58 & 1.33 & 0.99 & 98.79
& 49.80 & 24.70 & 75.70 & 62.40 & 0.37 & 32.62 \\

\midrule

\makecell[l]{\sysname\\w/o Alignment}
& 99.13 & 1.09 & 0.67 & 1.00 & 0.99 & 99.13
& 59.25 & 47.30 & 34.20 & 44.30 & 0.59 & 61.76 \\

\makecell[l]{\sysname\\w/o Subtraction}
& 98.83 & 1.25 & 1.08 & 1.25 & 0.99 & 98.84
& 53.25 & 39.90 & 53.60 & 43.10 & 0.56 & 49.81 \\

\makecell[l]{\sysname\\w/o Contrastive}
& 99.00 & 1.17 & 0.83 & 1.17 & 0.99 & 99.00
& 64.50 & 57.20 & 13.80 & 28.00 & 0.72 & 70.83 \\

\midrule

\textbf{\sysname}
& \textbf{99.38}
& \textbf{0.67}
& \textbf{0.58}
& \textbf{0.67}
& \textbf{0.99}
& \textbf{99.38}
& \textbf{93.25}
& \textbf{4.80}
& \textbf{8.70}
& \textbf{8.00}
& \textbf{0.96}
& \textbf{93.12} \\

\bottomrule
\end{tabular*}
\end{table*}
\subsubsection{Registered User Classification}
\label{registeduserclassification}

We first evaluate liveness classification for registered users. As shown in Table~\ref{tab:overall_performance}, \sysname achieves 99.38\% ACC, 0.67\% EER, 0.9996 AUC, and 99.38\% F1 score. MiniFAS performs worst because visual provides limited access to the intrinsic physical differences between genuine skin and 3D spoofing masks. Compared with AFace and MID, \sysname benefits from complementary electromagnetic and mechanical responses, yielding richer liveness evidence. Compared with mmFAS, \sysname further suppresses shared geometry variations, allowing the resulting representation to emphasize material-related liveness information and improve accuracy. Signal-based baselines perform well under material variations, because these variations are seen during training.

\subsubsection{Unseen User Classification}
\label{Unseen User Classification}
\begin{figure}[t]
    \centering
    \includegraphics[width=0.6\linewidth]{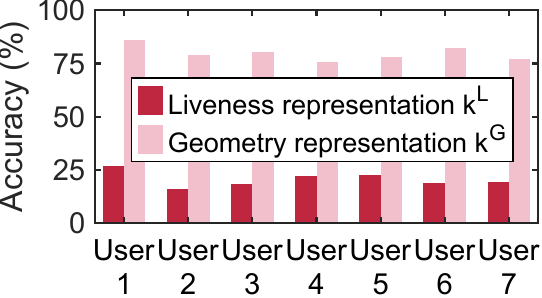}
    \caption{Validation of cross-modal subtraction.}
    \label{fig:disentanglement}
\end{figure}
We next evaluate generalization to unseen users following the protocol described in Sec.~\ref{subsec:experimental_setup}. As shown in Table~\ref{tab:overall_performance}, \sysname achieves 93.25\% ACC, 4.80\% FAR, 8.70\% FRR, 0.9695 AUC, and 93.12\% F1 score, substantially outperforming MiniFAS (50.40\%), AFace (50.50\%), MID (56.10\%), and mmFAS (49.80\%) in accuracy. The near-random performance of MiniFAS reflects the limited ability of visual appearance to distinguish high-fidelity masks. Although AFace and MID capture material-dependent physical responses, their representations remain affected by user-dependent facial geometry. Similarly, mmFAS does not remove shared geometry-related variations. In contrast, \sysname explicitly suppresses these user-dependent factors while retaining liveness-discriminative responses, resulting in substantially stronger cross-user generalization.

To verify whether cross-user generalization comes from suppressing user-dependent information, we conduct an identity-probing experiment on seven randomly selected users. After training \sysname, we freeze the feature extraction, alignment, and subtraction modules and train user classifiers on the liveness representation $\mathbf{k}^{L}$ and geometry representation $\mathbf{k}^{G}$. As shown in Fig.~\ref{fig:disentanglement}, $\mathbf{k}^{G}$ achieves 79.90\% user classification accuracy, whereas $\mathbf{k}^{L}$ achieves 20.48\%, closer to random guessing (14.29\%). This confirms that the additive branch retains substantial user-dependent information, while cross-modal subtraction suppresses it in the liveness representation.

\subsubsection{Ablation Study}
We further evaluate the contribution of the three key components. For the variants without alignment or subtraction, we retain the two-stage training strategy so that the contrastive-learning procedure remains unchanged; without contrastive projection, the remaining network is trained in a single stage. Under the registered-user setting, all ablated variants maintain strong performance across the evaluation metrics, as the model can still exploit user-specific patterns observed during training. In contrast, their roles become more evident for unseen users. As shown in Table~\ref{tab:overall_performance}, removing cross-modal alignment reduces unseen-user ACC from 93.25\% to 59.25\% and AUC from 0.9695 to 0.5954. Since AUC reflects live/spoof separability across decision thresholds, this sharp decrease indicates that semantic and scale mismatch makes subsequent subtraction unreliable and causes the two classes to become less separable. Removing subtraction causes the largest ACC degradation to 53.25\%, showing that without suppressing user-dependent geometry, the learned representation fails to generalize to unseen users. Without contrastive projection, ACC drops to 64.50\% and FAR rises from 4.80\% to 57.20\%. The substantial FAR increase indicates that material-induced feature drift causes spoofing presentations to be increasingly accepted as genuine even after geometry suppression. These results demonstrate that alignment enables reliable cross-modal comparison, subtraction supports cross-user generalization, and contrastive projection mitigates material-induced drift.
\begin{figure*}[t]
    \centering

    \begin{minipage}[t]{0.24\textwidth}
        \centering
        \includegraphics[width=\linewidth]{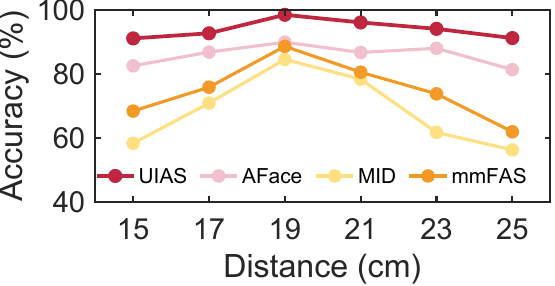}
        \captionof{figure}{Impact of sensing range.}
        \label{fig:impact_range}
    \end{minipage}%
    \begin{minipage}[t]{0.24\textwidth}
        \centering
        \includegraphics[width=\linewidth]{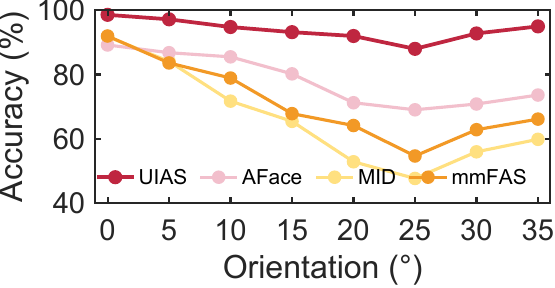}
        \captionof{figure}{Impact of head orientation.}
        \label{fig:impact_orientation}
    \end{minipage}%
    \begin{minipage}[t]{0.24\textwidth}
        \centering
        \includegraphics[width=\linewidth]{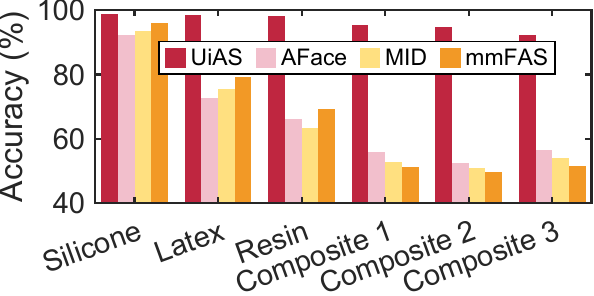}
        \captionof{figure}{Impact of unseen spoofing mask materials.}
        \label{fig:impact_material}
    \end{minipage}
     \begin{minipage}[t]{0.24\textwidth}
        \centering
        \includegraphics[width=\linewidth]{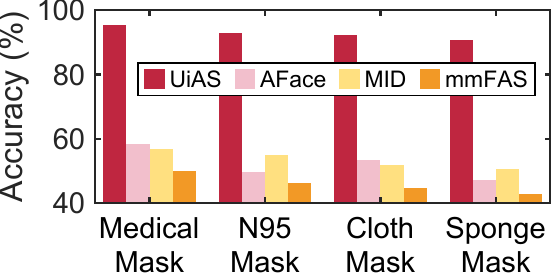}
        \captionof{figure}{Impact of unseen face-covering types.}
        \label{fig:impact_covering}
    \end{minipage}%

    \par

    \begin{minipage}[t]{0.24\textwidth}
        \centering
        \includegraphics[width=\linewidth]{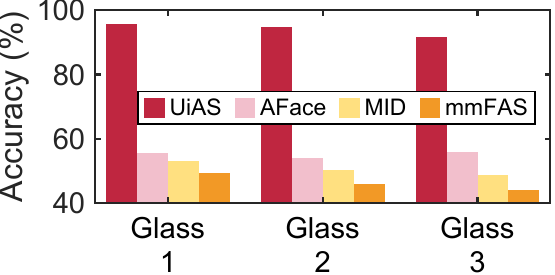}
        \captionof{figure}{Impact of unseen eyeglass.}
        \label{fig:impact_eyeglass}
    \end{minipage}%
    \begin{minipage}[t]{0.24\textwidth}
        \centering
        \includegraphics[width=\linewidth]{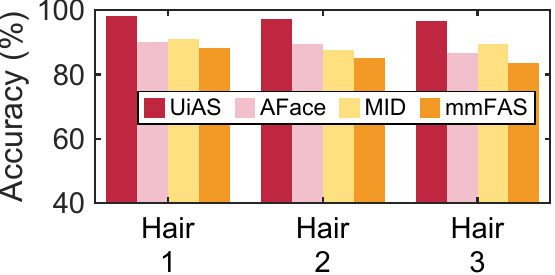}
        \captionof{figure}{Impact of unseen  hairstyles.}
        \label{fig:impact_hairstyle}
    \end{minipage}%
    \begin{minipage}[t]{0.24\textwidth}
        \centering
        \includegraphics[width=\linewidth]{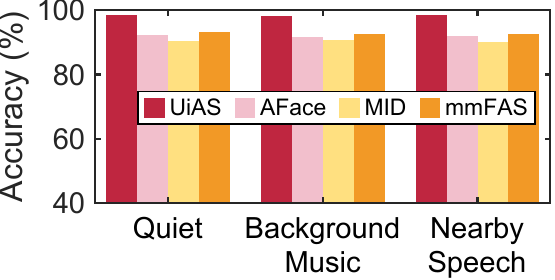}
        \captionof{figure}{Impact of noise.}
        \label{fig:impact_noise}
    \end{minipage}
    \begin{minipage}[t]{0.24\textwidth}
        \centering
        \includegraphics[width=\linewidth]{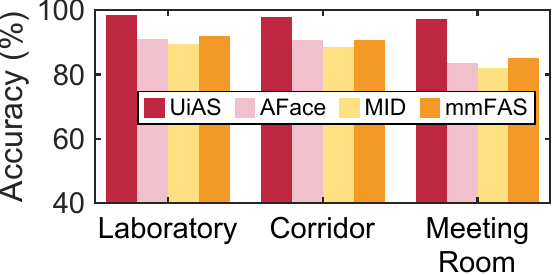}
        \captionof{figure}{Impact of environment.}
        \label{fig:impact_environment}
    \end{minipage}

\end{figure*}
\subsection{Impact of Factors}
\label{subsec:impact_factors}

To evaluate \sysname robust, we conduct controlled experiments over multiple factors and compare it with three wireless-based baselines: AFace, MID, and mmFAS. 
In each fold, we evaluate two seen and two unseen users to cover both user settings, and report results averaged across the seven folds.

\noindent\textbf{Impact of Sensing Geometry.}
\label{sssec:impact_geometry}
To evaluate robustness to sensing geometry, we vary the range from 15 to 25~cm at 2~cm intervals and head orientation from 0$^\circ$ to 35$^\circ$ at 5$^\circ$ intervals. Data acquisition is manually triggered after the participant reaches the designated sensing range and orientation. Each setting includes 250 genuine-face and 250 generic-mask trials per participant. As shown in Figs.~\ref{fig:impact_range} and~\ref{fig:impact_orientation}, \sysname maintains over 91.2\% accuracy across all ranges and 88.0\% across all orientations, whereas the baselines degrade more noticeably as the sensing range or head orientation changes. Because changes in sensing geometry affect both co-located modalities in a correlated manner, cross-modal subtraction suppresses not only user-dependent facial geometry but also shared variations induced by range and orientation, providing additional robustness to varying sensing geometries. Accuracy initially decreases and then partially recovers at larger orientations, likely because moderate rotations reduce the effective reflective area, whereas larger rotations expose lateral facial regions to the sensors and strengthen the received responses.

\noindent\textbf{Impact of Unseen Spoofing Mask Materials.}
\label{sssec:impact_mask_material}
To evaluate generalization to unseen spoofing materials, we test silicone masks and unseen latex and resin masks, and further attach moisturizing facial sheets to each mask type to introduce three composite conditions. Each participant completes 250 trials per condition. As shown in Fig.~\ref{fig:impact_material}, \sysname achieves at least 92.2\% accuracy across all unseen conditions. Existing methods degrade significantly because their decision boundaries are learned from spoof-material distributions observed during training. In particular, increased moisture content in composite masks makes some of their physical responses more skin-like, raising misclassification risk as genuine. In contrast, \sysname uses genuine skin as the reference and evaluates whether the projected liveness representation is consistent with the stable skin characteristics represented by the skin anchor. Although the composite materials become more skin-like in some physical responses, they still cannot reproduce the genuine-skin characteristics captured by this reference, enabling \sysname to remain robust to such unseen materials.

\noindent\textbf{Impact of Unseen Material Variations.}
\label{sssec:impact_multimaterial}
To evaluate robustness to unseen material variations, we consider face coverings, eyeglasses, and hairstyles, where the specific materials used for testing are not observed during training.
For each variation condition, each participant completes 250 genuine-face and 250 generic-mask trials. As shown in Figs.~\ref{fig:impact_covering}, \ref{fig:impact_eyeglass}, and~\ref{fig:impact_hairstyle}, \sysname consistently outperforms the baselines, achieving at least 90.7\% accuracy for face coverings, 91.8\% for eyeglasses, and 96.6\% for hairstyles. Existing methods degrade because these variations introduce additional physical responses that shift the liveness representation. In contrast, the skin-anchored projection in \sysname maps material-affected representations into a skin-referenced liveness space, suppressing variations inconsistent with genuine-skin characteristics while preserving skin-related liveness evidence. Hairstyle variations cause milder degradation because hair covers a smaller facial region and introduces weaker reflection changes than the others.

\noindent\textbf{Impact of Noise.}
\label{sssec:impact_noise}
We evaluate all methods under quiet, background-music, and nearby-speech conditions, with 250 genuine-face and 250 generic-mask trials per participant in each setting. As shown in Fig.~\ref{fig:impact_noise}, all methods remain relatively stable because the acoustic-based methods operate mainly at high frequencies, while MID uses only mmWave sensing. Thus, common audible noise has limited impact.

\noindent\textbf{Impact of Environments.}
We evaluate all methods in a laboratory, corridor, and meeting room, with 250 genuine-face and 250 generic-mask trials per participant in each environment. As shown in Fig.~\ref{fig:impact_environment}, the baselines degrade in the meeting room due to stronger multipath reflections, while \sysname remains stable because target-region preprocessing and suppresses environmental multipath interference.

\noindent\textbf{Computation Overhead.}
We evaluate the runtime overhead of \sysname on an NVIDIA RTX 4090 GPU. The average inference latency is 0.11~ms per sample, substantially lower than the sensing interval, supporting real-time liveness detection.
\section{Related Work}
\label{sec:related}

\noindent\textbf{2D Face Anti-spoofing.}
Vision-based face anti-spoofing (FAS) has been widely studied for detecting 2D presentation attacks, such as printed photos and replayed videos~\cite{cvprinted}. Existing methods exploit texture~\cite{cvtexture},  motion~\cite{cvmotion1,cvmotion2,cvmotion3} and reflection pattern~\cite{faceflash} from facial images and videos. To improve robustness, multimodal FAS systems further combine complementary channels, such as depth~\cite{cvdepth1,cvdepth2,mmfas}, infrared~\cite{cvmultimodal1,cvmultimodal2,cvmultimodal3} and ultrasonic sensing~\cite{echoprint,m3fas}. For example, RGB--thermal methods fuse visible appearance with thermal emission~\cite{rgbthermal}, while SONICUMOS~\cite{sonicumos} combines video and ultrasound to verify active head gestures. However, these methods are primarily designed or evaluated for planar media attacks. High-fidelity 3D masks can preserve realistic appearance, facial contours, and depth structures~\cite{cvmask}, making 2D-oriented FAS insufficient for 3D spoofing attacks.

\noindent\textbf{3D Face Anti-spoofing.}
To address 3D attacks, prior work has explored face anti-spoofing using RF or acoustic signals. RFace~\cite{rface} extracts facial geometry and material responses from RFID reflections, while mmFace~\cite{mmface} uses mmWave synthetic-aperture radar (SAR) sensing to recover facial characteristics, requiring spatial scanning that limits practical deployment. Acoustic systems characterize facial structures or dynamic motions~\cite{echoface,beyond,sonarguard}, while recent systems such as AFace~\cite{aface} and UltraFace~\cite{ultraface} exploit range-adaptive or impedance-related responses. These approaches capture richer physical evidence than RGB. However, RF and acoustic responses jointly encode liveness cues and user-dependent facial geometry, becoming entangled in learned representations. Existing systems rely on user-specific enrollment or template-based verification without suppressing geometry-related variations, limiting generalization to unseen users.

\noindent\textbf{Multi-modal Learning and Sensing.}
Multimodal learning and sensing integrates complementary information across heterogeneous modalities~\cite{clip,radarcam,rcbevdet,cosmo,hydra,cost,proteus}. Existing methods typically concatenate features, apply attention-based fusion, or align modalities in a shared space~\cite{align}. However, in bimodal facial sensing, both modalities encode user-dependent information, so naive fusion or alignment may preserve these nuisance factors rather than isolate generalizable liveness cues. Moreover, existing disentanglement methods often assume balanced or identifiable factor contributions~\cite{decoupling1,decoupling2}, whereas liveness cues are much weaker than geometry variations. Therefore, user-independent 3D face anti-spoofing requires suppressing shared user-dependent information instead of merely aggregating multimodal features.

\sssec{$\blacksquare$ Summary of Differences.}
\sysname enables user-independent 3D face anti-spoofing with mmWave--acoustic sensing by suppressing shared user-dependent variations while preserving liveness responses, eliminating user-specific enrollment. 
\section{Discussion and Limitations}
\label{sec:discussion}

\noindent\textbf{Integration with facial recognition systems.}
\sysname is designed as a plug-in verification module for existing face recognition pipelines without modifying the recognition model or redesigning the authentication circuitry. Moreover, some emerging smart devices already integrate compact mmWave and acoustic components~\cite{acuse,mmuse}. Such hardware can potentially be reused to miniaturize \sysname and incorporate physical-response verification into existing authentication systems.

\noindent\textbf{Practicality of multi-material presentations.}
\sysname tolerates common materials such as eyeglasses and face coverings during facial anti-spoofing. However, in practical high-security scenarios, such materials can be required to be manually removed, such as mobile banking, border and etc.

\noindent\textbf{Wireless injection and sensor dependency.}
An attacker could potentially inject RF or acoustic signals within sensing bands, contaminating the responses and potentially affecting liveness decisions. Moreover, in large-scale deployment, sensors from different manufacturers or models may inevitably introduce distribution shifts. However, \sysname focuses on the mechanism for user-independent liveness detection, while signal-level injection defenses and cross-device calibration are beyond the scope of this paper and left for future work.

\noindent\textbf{From single-target to multi-target anti-spoofing.}
\sysname currently targets individual detection, where the user approaches the sensing system and dominates the captured response. In crowded scenarios, responses from multiple users may overlap across modalities, making target association challenging. Future work can incorporate larger arrays, metasurfaces~\cite{metasurface}, beamforming~\cite{beamforming}, and tracking for multi-target anti-spoofing.

\section{Conclusion}
This paper presents \sysname, a user-independent 3D facial anti-spoofing system using co-located mmWave and acoustic sensing. \sysname suppresses cross-modal shared user-dependent variations while preserving liveness-discriminative physical responses. Experiments on 14 subjects show that \sysname achieves 93.25\% detection accuracy and outperforms baselines under unseen-user settings. These results demonstrate the effectiveness of cross-modal nuisance suppression for user-independent 3D face anti-spoofing. 
\cleardoublepage
\appendix
\section*{Ethical Considerations}
\label{app:ethics}

\noindent\textbf{Human participants and informed consent.}
Our study involves human participants and the collection of
face-related sensing measurements, including mmWave and acoustic
responses, as well as camera information used to define the facial
sensing region. Participants span three self-identified racial groups (Asian, White, and Black). All participants were informed of the purpose and
procedure of the study and voluntarily provided informed consent
before data collection. Participants whose images appear in the
paper additionally consented to the use of these images for research
publication.

\noindent\textbf{Participant privacy.}
The collected measurements encode participant-specific facial
characteristics and are therefore treated as privacy-sensitive
biometric information. Access to the collected data is restricted
to the research team. We do not publicly release either the raw or
processed participant data, including facial images, mmWave and
acoustic facial measurements, or participant-specific 3D facial
models, because these data may reveal biometric characteristics of
the participants.

\noindent\textbf{Spoofing experiments.}
The high-fidelity 3D masks and spoofing attacks evaluated in this
work were created and tested exclusively in controlled research
settings. Experiments involving commercial face-authentication
systems were performed only on devices and accounts under the
researchers' control. No third-party accounts, devices, or personal
data were accessed. We anonymize the tested commercial device
models to avoid unnecessarily exposing device-specific attack
information.

\noindent\textbf{Potential misuse.}
The techniques and experimental results presented in this work may
provide information relevant to facial spoofing attacks. Our purpose
is to characterize this threat and develop defenses against it.
To reduce potential misuse and protect participants, we do not
release participant-specific 3D facial models or biometric sensing
measurements. We release only the implementation of the proposed
defensive system.
\cleardoublepage
\section*{Open Science} \label{app:open_science}
To support reproducibility and future research, we provide the following artifact and data availability statements.

\noindent\textbf{Artifact availability.} An anonymized version of the artifact is available at: 
\begin{center} \url{https://anonymous.4open.science/r/SEC27-artifact-UiAs} \end{center} 
The artifact contains the implementations necessary to reproduce the signal processing, model architecture, training objectives, evaluation pipeline, and demonstration video described in this paper. Participant-derived data are excluded from the artifact for privacy protection. 

\noindent\textbf{Data availability.} We do not release either the raw or processed participant data collected in this study. The dataset contains biometric information derived from participants' facial characteristics, including mmWave and acoustic facial responses and participant-related visual information. Even after preprocessing, these measurements remain associated with individual facial characteristics and may introduce privacy risks if publicly released. Therefore, consistent with the participant privacy and data-management constraints of this study, the dataset is not included in the public artifact.
\cleardoublepage

\bibliographystyle{plainurl}
\bibliography{ref}

\end{document}